\documentclass[twocolumn,superscriptaddress]{revtex4-1}
\usepackage{color}
\usepackage{graphics,graphicx,epsfig}
\usepackage{epsf,epstopdf,wrapfig}
\usepackage{amssymb,amsfonts,amsmath}
\usepackage[export]{adjustbox}
\include{epsf}
\usepackage{ifthen}		

\newcommand{\beginsupplement}{%
        \setcounter{table}{0}
        \renewcommand{\thetable}{S\arabic{table}}%
        \setcounter{figure}{0}
        \renewcommand{\thefigure}{S\arabic{figure}}%
     }

\newcommand{\beqn}{\begin{eqnarray}}
\newcommand{\eeqn}{\end{eqnarray}}
\newcommand{\beq}{\begin{equation}}
\newcommand{\eeq}{\end{equation}}

\newcommand{\abs}[1]{|#1|}
\newcommand{\<}{\langle}
\renewcommand{\>}{\rangle}

\newcommand{\bs}{ \mbox{\boldmath$\sigma$}}

\newcommand{\ssection}[1]{\subsection{#1}}
\newcommand{\mref}[1]{Sec.~\ref{#1}}

\definecolor{junglegreen}{rgb}{0.16, 0.67, 0.53}
\definecolor{myrtle}{rgb}{0.13, 0.26, 0.12}
\definecolor{lincolngreen}{rgb}{0.11, 0.35, 0.02}
\definecolor{forestgreen}{rgb}{0.13, 0.55, 0.13}

\newcommand{\rev}[1]{{#1}}

\newcommand{\lastequal}{These authors contributed equally. Please
  correspondance to \url{awalczak@lpt.ens.fr}, \url{tmora@lps.ens.fr}}

\graphicspath{{final_plots/}}
\begin{document}
\title{Size and structure of the sequence space of repeat proteins}

\author{Jacopo Marchi}
\affiliation{Laboratoire de physique de l'\'Ecole normale sup\'erieure (PSL University), CNRS, Sorbonne Universit\'e, and Universit\'e de Paris, 75005 Paris, France}
\author{Ezequiel A. Galpern}
\affiliation{Protein Physiology Lab, Universidad de Buenos Aires, Facultad de Ciencias Exactas y Naturales, Departamento de Qu\'\i mica Biol\'ogica. Buenos Aires, Argentina. / CONICET - Universidad de Buenos Aires. Instituto de Qu\'\i mica Biol\'ogica de la Facultad de Ciencias Exactas y Naturales (IQUIBICEN). Buenos Aires, Argentina}
\author{Rocio Espada}
\affiliation{Laboratoire Gulliver, Ecole sup\'erieure de physique et chimie industrielles (PSL University) and CNRS, 75005, Paris, France}
\author{Diego U. Ferreiro}
\affiliation{Protein Physiology Lab, Universidad de Buenos Aires, Facultad de Ciencias Exactas y Naturales, Departamento de Qu\'\i mica Biol\'ogica. Buenos Aires, Argentina. / CONICET - Universidad de Buenos Aires. Instituto de Qu\'\i mica Biol\'ogica de la Facultad de Ciencias Exactas y Naturales (IQUIBICEN). Buenos Aires, Argentina}
\author{Aleksandra M. Walczak}
\affiliation{Laboratoire de physique de l'\'Ecole normale sup\'erieure (PSL University), CNRS, Sorbonne Universit\'e, and Universit\'e de Paris, 75005 Paris, France}
\affiliation{\lastequal}
\author{Thierry Mora}
\affiliation{Laboratoire de physique de l'\'Ecole normale sup\'erieure
  (PSL University), CNRS, Sorbonne Universit\'e, and Universit\'e de
  Paris, 75005 Paris, France}
\affiliation{\lastequal}

\begin{abstract}
 % abstract
The coding space of protein sequences is shaped by evolutionary constraints set by requirements of function and stability. We show that the coding space of a given protein family ---the total number of sequences in that family--- can be estimated using models of maximum entropy trained on multiple sequence alignments of naturally occuring amino acid sequences. We analyzed and calculated the size of three abundant repeat proteins families, whose members are large proteins made of many repetitions of conserved portions of $\sim 30$ amino acids.
While amino acid conservation at each position of the alignment explains most of the reduction of diversity relative to completely random sequences,
we found that correlations between amino acid usage at different positions significantly impact that diversity. We quantified the impact of different types of correlations, functional and evolutionary, on sequence diversity. Analysis of the detailed structure of the coding space of the families revealed a rugged landscape, with many local energy minima of varying sizes with a hierarchical structure, reminiscent of fustrated energy landscapes of spin glass in physics. This clustered structure indicates a multiplicity of subtypes within each family, and suggests new strategies for protein design.

\end{abstract}

\maketitle

\section{Introduction}

 % intro

Natural proteins contain a record of their evolutionary history, as selective pressure constrains their amino-acid sequences to perform certain functions. However, if we take all proteins found in nature, their sequence appears to be random, without any apparent rules that distinguish their sequences from arbitrary polypeptides. Nonetheless, the volume of sequence space taken up by existing proteins is very small compared to all possible polypeptide strings of a given length~\cite{Dryden2008}, \rev{even more so when specializing to a given structure \cite{Shakhnovich1998}}. Clearly, not all variants are equally likely to survive~\cite{Salisbury1969, Mandecki1998, Dokholyan2002}. To better understand the structure of the space of natural proteins, it is useful to group them into families of proteins with similar fold, function, and sequence, believed to be under a common selective pressure. Assuming that the ensemble of protein families is equilibrated, there should exist a relationship between the conserved features of their amino acid sequences and their function. This relation can be extracted by examining statistics of amino-acid composition, starting with single sites in multiple alignments (as provided by e.g. PFAM~\cite{bateman2004pfam,finn2013pfam}). More interesting information can be extracted from covariation of amino acid usages at pairs of positions \cite{neher1994frequent,pmid24573474,Szurmant2018} or using machine-learning techniques~\cite{Tubiana2018}. Models of protein sequences based of pairwise covariations have been shown to successfully predict pair-wise amino-acid contacts in three dimensional structures~\cite{weigt2009pnas,morcos2011pnas,aurell2013pre, hopf2012three, espada2015BMC, Espada2018}, aid protein folding algorithms~\cite{Schug2009,marks2011protein}, and predict the effect of point mutations \cite{Espada2018,Contini, Haldane2016,weigt2015mbe}.
However, little is known on how these identified amino-acid constraints affect the global size, shape and structure of the sequence space. Accounting for these questions is a first step towards drawing out the possible and the realized evolutionary trajectories of protein sequences ~\cite{MaynardSmith1970,Weinreich2006}.

We use tools and concepts from the statistical mechanics of disordered systems to study collective, protein-wide effects and to understand how evolutionary constraints shape the landscape of protein families. We go beyond previous work which focused on local effects
--- pairwise contacts between residues, effect of single amino-acid mutations --- to ask how amino-acid conservation and covariation {restrict} and {shape} the landscape of sequences in a family. Specifically, we characterize the size of the ensemble, defined as the effective number of sequences of a familiy, as well as its detailed structure: is it made of one block or divided into clusters of ``basins''? These are intrinsically collective properties that can not be assessed locally.

\begin{figure}
\includegraphics[width=.7\linewidth]{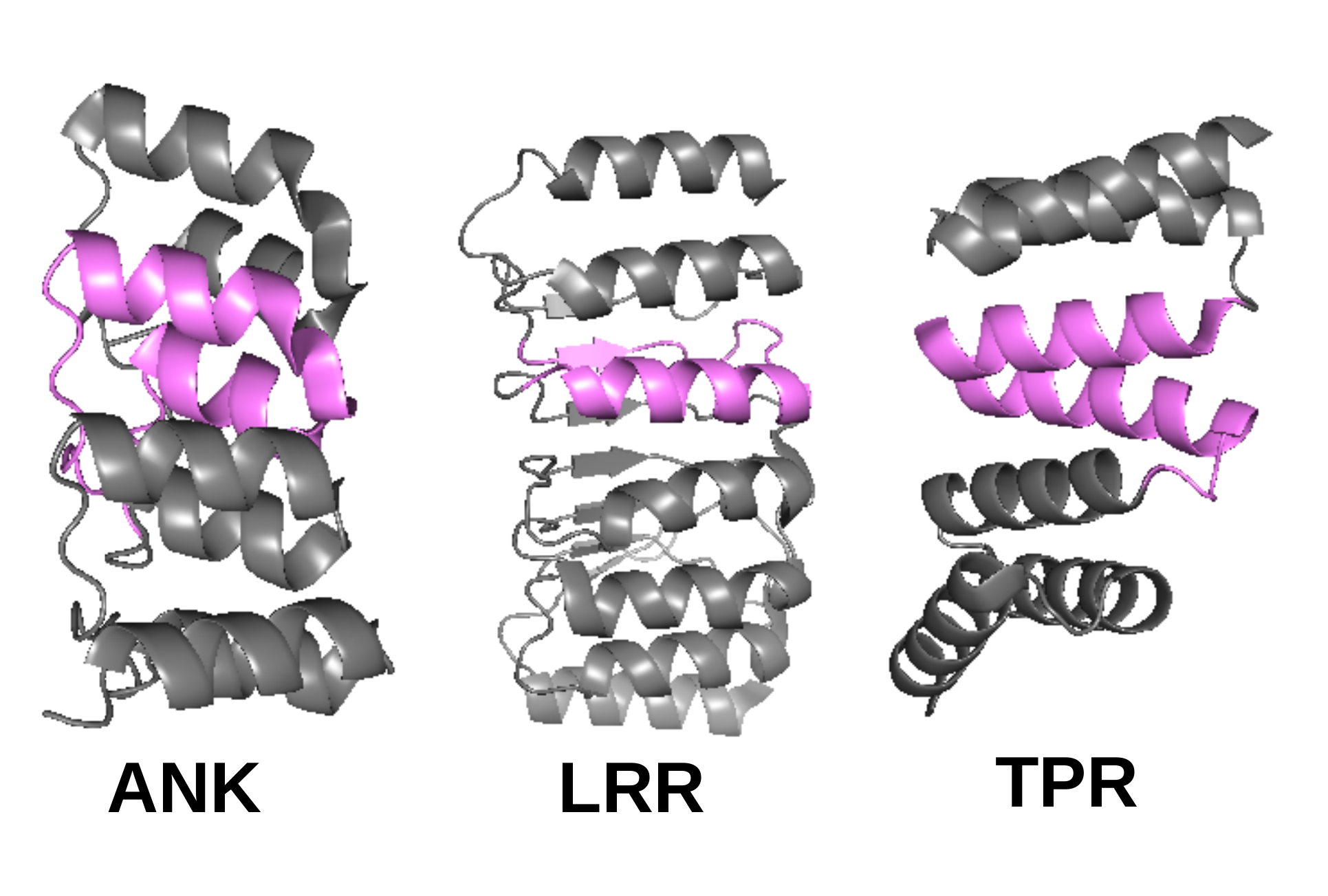}
\caption{Repeat proteins fold into characteristic accordeon-like folds. Example structures of three protein families are shown, ankyrin repeats (ANK), tetratricopeptide repeats (TPR), leucine-rich repeat (LRR), with the repeating unit highlighted in magenta. All show regular folding patterns with defined contacts in and between repeats.  \label{Fig0}}
\end{figure}

Repeat proteins are excellent systems in which to quantify these collective effects, as they combine both local and global interactions. Repeat proteins are found as domains or subdomains in a very large number of functionally important proteins, in particular signaling proteins (e.g. NF-$\kappa$B, p16, Notch~\cite{pmid17176038}). Usually they are composed of tandem repetitions of $\sim30$ amino-acids that fold into elongated architectures.
Repeat proteins have been divided into different families based on their structural similarity. Here we consider three abundant repeat protein families: ankyrin repeats (ANK), tetratricopeptide repeats (TPR), leucine-rich repeat (LRR) that fold into repetitive structures (see Fig.~\ref{Fig0}). In addition to interactions between residues within one repeat, repeat protein evolution is constrained by inter-repeat interactions, which lead to the characteristic accordeon-like folds. Through these separable types of constraints, as well as the possibility of intra- and inter-familly comparisons, repeat proteins are perfect candidates to ask questions about the origins and the effects of the constraints that globally shape the sequences. 

A recent study~\cite{Tian2017} addressed the question of the total number of sequences within a given protein family, focusing on ten single-domain families. They took a similar thermodynamic approach to the one followed here, but had to estimate experimentally the free energy threshold $\Delta G$ below which the sequences would fold properly. Here we overcome this limitation by forgoing this threshold entirely. Instead we determine the sequence entropy directly, which is argued to be equivalent to using a threshold free energy by virtue of the equivalence of ensembles. We precisely quantify the sequence entropy of three repeat-protein families for which detailed evolutionary energetic fields are known~\cite{Barton2016a}. We explore the properties of the evolutionary landscape shaped by the amino-acid frequency constraints and correlations. We ask whether the energy landscape, defined in sequence space of repeat proteins, is made of a single basin, or rather of a multitude of basins connected by ridges and passes, called ``metastable states'', as would be expected from spin-glass theory.
Using the specific example of repeat proteins makes it possible to analyze the source of the potential landscape ruggedness, and use it to identify which repeat-protein families can be well separated into subfamilies. The rich metastable state structure that we find demonstrates the importance of interactions in shaping the protein family ensemble.

\section{Results}

 % Potts_fitting
\ssection{Statistical models of repeat-protein families}\label{Potts_fit}
We start by building statistical models for the three repeat protein families presented in Fig.~\ref{Fig0} (ANK, TPR, LRR). These models give the probability $P(\bs)$ to find in the family of interest a particular sequence $\bs=(\sigma_1,\ldots,\sigma_{2L})$ for two consecutive repeats of size $L$. The model is designed to be as random as possible, while agreeing with key statistics of variation and co-variation in a multiple sequence alignment of the protein family. Specifically, $P(\bs)$ is obtained as the distribution of maximum entropy \cite{Jaynes} which has the same amino-acid frequencies at each position as in the alignment, as well as the same joint frequencies of amino acid usage in each pair of positions. Additionally, repeat proteins share many amino acids between consecutive repeats, both due to sharing a common ancestor and to evolutionary selection acting on the protein. To account for this special property of repeat proteins, we require that the model reproduces the distribution of overlaps \rev{${\rm ID}(\bs)=\sum_{i=1}^L\delta_{\sigma_i,\sigma_{i+L}}$ between consecutive repeats}. Using the technique of Lagrange multipliers, the distribution can be shown to take the form~\cite{Espada2018}:
\beq
P(\bs)=(1/Z)e^{-E(\bs)}, \label{eq:boltzmann}
\eeq
with
\begin{equation}\label{H_ro} 
E(\bs)= - \sum_{i=1}^{2L} h_i(\sigma_i) -\sum_{i,j=1}^{2L} J_{ij}(\sigma_i, \sigma_j)  + \lambda_{{\rm ID}}(\bs) \,,
\end{equation}
\rev{where $h_i(\sigma)$, $J_{ij}(\sigma_i, \sigma_j)$, and $\{\lambda_{\rm ID}\}$, ${\rm ID}=0,1,\ldots,L$, are adjustable Lagrange multipliers that are fit to the data to reproduce the experimentally observed site-dependent amino-acid  frequencies $f_i(\sigma_i)$, joint probabilities between two positions, $f_{ij}(\sigma_i, \sigma_j)$, and the distribution of Hamming distances between consecutive repeats $P({\rm ID}(\bs))$, which is equivalent to maximize the likelihood of the data under the model.
We fit these parameters using a gradient ascent algorithm: we start from an initial guess of the parameters, then generate  sequences via Monte-Carlo simulations and update the parameters proportionally to the difference between the empirical and model generated observables $f_i(\sigma_i) - f_i^{\rm model}(\sigma_i)$, $f_{ij}(\sigma_i, \sigma_j) - f_{ij}^{\rm model}(\sigma_i, \sigma_j)$ and $P({\rm ID}(\bs)) - P({\rm ID}(\bs))^{\rm model}$. We repeat the previous steps 
until the model reproduces the empirical observables defined above, with a target precision motivated according to the finite size of our original dataset,  as in Ref.~\cite{Espada2018}.
 See \mref{model_fitting} for more details. We tested the convergence of the model learning by synthetically generating datasets and relearning the model (see \mref{entropy_error}).}

By analogy with Boltmzan's law, we call $E(\bs)$ a statistical energy, which is in general distinct from any physical energy. The particular form of the energy \eqref{H_ro} resembles that of a disordered Potts model. This mathematical equivalence allows for the possibility to study effects that are characteristic of disordered systems, such as frustration or the existence of an energy landscape with multiple valleys, as we will discuss in the next sections.

Eq.~\ref{H_ro} is the most constrained form of the model, which we will denote by $E_{\rm full}(\bs)$.
One can explore the impact of each constraint on the energy landscape by removing them from the model.
For instance, to study the role of inter-repeat sequence similarity due to a common evolutionary origin, one can fit the model without the constraint on repeat overlap ID, {\em i.e.} without the $\lambda_{{\rm ID}}$ term in Eq.~\ref{H_ro}. We call the corresponding energy function $E_{\rm 2}$. One can further remove constraints on pairwise positions that are not part of the same repeat, making the two consecutive repeats statistically independent and imposing $h_i=h_{i+L}$ ($E_{\rm ir}$), or only linked through phylogenic conservation through $\lambda_{\rm ID}$ ($E_{\rm ir,\lambda}$). Finally one can remove all interaction constraints  to make all positions independent of each other ($E_{\rm 1}$), or even remove all constraints ($E_{\rm rand}\equiv 0$).

 % deltaG_Potts
\ssection{Statistical energy {\em vs} unfolding energy}
The evolutionary information contained in multiple sequence alignments of protein families is summarized in our model by the energy function $E(\bs)$. Since this information is often much easier to access than structural or functional information, there is great interest in extracting functional or structural properties from multiple sequence alignments, provided that there exists a clear quantitative relationship between statistical energy and physical energy.

Such a relationship was determined experimentally for repeat proteins
by using $E(\bs)$ to predict the effect of point mutations on the
folding stability measured by the free energy difference between the
folded and unfolded states, $\Delta G$, called the unfolding
energy~\cite{Espada2018,Contini}. \rev{Synthetic sequences with low $E(\bs)$ have also been shown to reproduce the fold and function of natural sequences \cite{Tian2018}.}
Here, extending an argument already
developed in previous work \cite{Shakhnovich1993,Shakhnovich1993b,Dokholyan2001,Morcos2014}, we show how this correspondance between statistical likelihood and folding stability arises in a simple model of evolution.

Evolutionary theory predicts that the prevalence of a particular
genotype $\bs$, i.e. the probability of finding it in a population, is related
to its fitness $F(\bs)$. 
In the limit where mutations affecting the protein are rare compared to the time it takes for mutations to spread through the population,
Kimura \cite{Kimura1962} showed that the probability of a mutation
giving a fitness advantage (or disadvantage depending on the sign)
$\Delta F$ over its ancestor will fix in the population with
probability $2\Delta F/(1-e^{-2N\Delta F})$, where $N$ is the effective
population size. The dynamics of successful
substitution satisfies detailed balance \cite{Berg2004}, with the steady state
probability
\beq\label{eq:kimura}
P(\bs)=(1/Z) e^{2NF(\bs)}.
\eeq
Again, one may recognize a
formal analogy with Boltzmann's distribution, where $F$ plays the role
of a negative energy, and $N$ an inverse temperature.
If we now assume that fitness is determined by the unfolding free energy
$\Delta G$, $F(\bs)=f(\Delta G(\bs))$, then
the distribution of genotypes we expect to observe in a population is
\beq
P(\bs)=(1/Z)e^{2Nf(\Delta G(\bs))}.
\eeq
Note that a similar relation should hold even if we relax the hypotheses of the evolutionary model. While in more general contexts (e.g. high mutation rate, recombination), the relation between $\ln P(\bs)$ and $F(\bs)$ may not be linear, such nonlinearities could be subsumed into the function $f$.

Identifying
terms in the two expressions (\ref{eq:boltzmann}) and (\ref{eq:kimura}), we obtain a relation between
the statistical energy $E$, and the unfolding free energy $\Delta G$:
\beq
E(\bs)=-2Nf(\Delta G(\bs)).
\eeq
For instance, if we assume a linear relation between fitness and
$\Delta G$, $f(\Delta G)= A+B \Delta G$, then we get a linear
relationship between the statistical energy and $\Delta G$, as was found empirically for repeat proteins \cite{Espada2018}.

Strikingly, the relationship $f$ does not have to be linear or even smooth for this correspondance to work.
Imagine a more stringent selection model, where $f(\Delta G)$
is a threshold function, $f(\Delta G)=0$ for
$\Delta G>\Delta G_{\rm sel}$ and $-\infty$
otherwise (lethal). In that case the probability
distribution is $P(\bs)=(1/Z)\Theta(\Delta G-\Delta G_{\rm sel})$,
where $\Theta(x)$ is Heaviside's function. Using a saddle-point approximation, one can show that in the
thermodynamic limit (long proteins, or large $L$) the distribution concentrates at the border $\Delta G_{\rm sel}$, and is equivalent to a ``canonical''
description~\cite{Shakhnovich1993,Shakhnovich1993b,Morcos2014}:
\beq\label{eq:sel}
P_{\rm sel}(\bs)=(1/Z)e^{\Delta G(\bs)/T_{\rm sel}},
\eeq
where the ``temperature'' $T_{\rm sel}$ is set to match the mean $\Delta G$ between the two descriptions:
\beq\label{eq:condition}
\<\Delta G\>_{T_{\rm sel}}=\Delta G_{\rm sel}.
\eeq
This correspondance is mathematically similar to the equivalence between the micro-canonical and canonical ensembles in statistical mechanics.

Statistical energy and unfolding free energy
are linearly related by equating (Eq.~\ref{eq:boltzmann}) and (Eq.~\ref{eq:sel}):
\beq\label{eq:EvsG}
E(\bs)=E_0-\Delta G(\bs)/T_{\rm sel},
\eeq
despite $f$ being nonlinear. Eq.~\ref{eq:EvsG}
is in fact very general and should hold for any $f$ in the thermodynamic limit in the vicinity of $\<E\>$.

 % entropy_calc
\ssection{Equivalence between two definitions of entropies}
\begin{table*}
\centering
\begin{tabular}{l*{7}{|c}}
family              & $2L$ & $S_{\rm rand}$ & $S_{1}$ & $S_{2}$ & $S_{\rm full}$  & $S_{\rm ir}$ & $S_{\rm ir,\lambda}$  \\
\hline
ANK              & 66 & 290 & 181  $\pm$ 0.05 &  169.7 $\pm$ 0.6  & 167.2 $\pm$ 0.3  & 176.7 $\pm$ 0.1& 172  $\pm$ 0.4   \\
LRR            & 48& 211 & 130 $\pm$ 0.05 &  114  $\pm$ 0.4  & 113.2  $\pm$ 0.3  & 123.1  $\pm$ 0.1 & 118.8  $\pm$ 0.1     \\
TPR 		& 68 & 299 & 169 $\pm$ 0.1 & 145.4  $\pm$ 0.7 & 141.4  $\pm$ 0.3 & 157.6  $\pm$ 0.1  & 146.9  $\pm$ 0.4    \\
\end{tabular}
\caption{Entropies (in bits, i.e. units of $\ln(2)$) of sequences made of two consecutive repeats, for the three protein families shown in Fig.~\ref{Fig0}. Entropies are calculated for models of different complexity: model of random amino acids ($S_{\rm rand}=2L\ln(21)$, divided by $\ln(2)$ when expressed in bits); independent-site model ($S_{1}$), pairwise interaction model ($S_{2}$); pairwise interaction model with constraints due to repeat similarity $\lambda_{\rm ID}$ ($S_{\rm full}$); pairwise interaction model of two non-interacting repeats learned without ($S_{\rm ir}$) and with ($S_{\rm ir,\lambda}$) constraints on repeat similarity. Fig.~\ref{thefig} shows graphically some of the information contained in this table.}
\label{table:entropies}
\end{table*}

There are several ways to define the diversity of a protein family. The most intuitive one, followed by \cite{Tian2017}, is to count the total number of amino acid sequences that have an unfolding free energy $\Delta G_{\rm sel}$ above a threshold $\Delta G_{\rm sel}$ \cite{Shakhnovich1998}. This number naturally defines a Boltzmann entropy,
\beq\label{eq:boltzmannentropy}
S=\ln \mathcal{N}(\bs: \Delta G(\bs)>\Delta G_{\rm sel}).
\eeq
Alternatively, starting from a statistical model $P(\bs)$, one can calculate its Shannon entropy, defined as
\beq\label{eq:shannon}
S=-\sum_{\bs} P(\bs)\ln P(\bs),
\eeq
as was done in Ref.~\cite{Barton2016a}. What is the relation between these two definitions?

By the same saddle-point approximation as in the previous section, the two are identical in the thermodyamic limit (large $L$), provided that the condition (Eq.~\ref{eq:condition}) is
satisfied. We can thus reconcile the two definitions of the entropy in that limit.

To calculate the Boltzmann entropy (Eq.~\ref{eq:boltzmannentropy}), one needs to first evaluate the threshold $E_{\rm sel}$ in terms of statistical energy. This threshold is given by $E_{\rm sel}=E_0- \Delta G_{\rm sel}/T_{\rm sel}$, where $E_0$ and $T_{\rm sel}$ can be obtained directly by fitting (Eq.~\ref{eq:EvsG}) to single-mutant experiments.
$E_{\rm sel}$ can also be obtained as a discrimination threshold separating sequences that are known to fold properly versus sequences that do not \cite{Tian2017}. In that case, assuming that the linear relationship (Eq.~\ref{eq:EvsG}) was evaluated empirically using single mutants, this relationship can be inverted to get $\Delta G_{\rm sel}$ in physical units.

Calculating the Shannon entropy Eq.~\eqref{eq:shannon}, on the other hand, does not require to define any threshold. However, the threshold in the equivalent Boltzmann entropy can be obtained using Eqs.~\ref{eq:condition} and \ref{eq:EvsG}, i.e. $E_{\rm sel}=\<E\>$, where the average is performed using the distribution defined in Eqs.~\ref{eq:boltzmann}-\ref{H_ro}.

\ssection{Entropy of repeat protein families}

To compare how the different elements of the energy function affect diversity, we calculate the entropy of ensembles built of two consecutive repeats from a given protein family for the different kinds of models described earlier, from the least constrained to the most constrained: $E_{\rm rand}$, $E_1$, $E_{\rm ir}$, $E_{\rm ir,\lambda}$, $E_2$, $E_{\rm full}$. In the case of models with interactions, calculating the entropy directly from the definition Eq.~\eqref{eq:shannon} is impossible due to the large sums. A previous study of entropies of protein families used an approximate mean-field algorithm, called the Adaptive Cluster Expansion \cite{Barton2016a}, for both parameter fitting and entropy estimation. Here we estimated the entropies using thermodynamic integration of Monte-Carlo simulations, as detailed in \mref{entropy_calc_meth:sec}. This method is expected to be asymptotically unbiased and accurate in the limit of large Monte-Carlo samples.

\begin{figure}
\includegraphics[width=\linewidth]{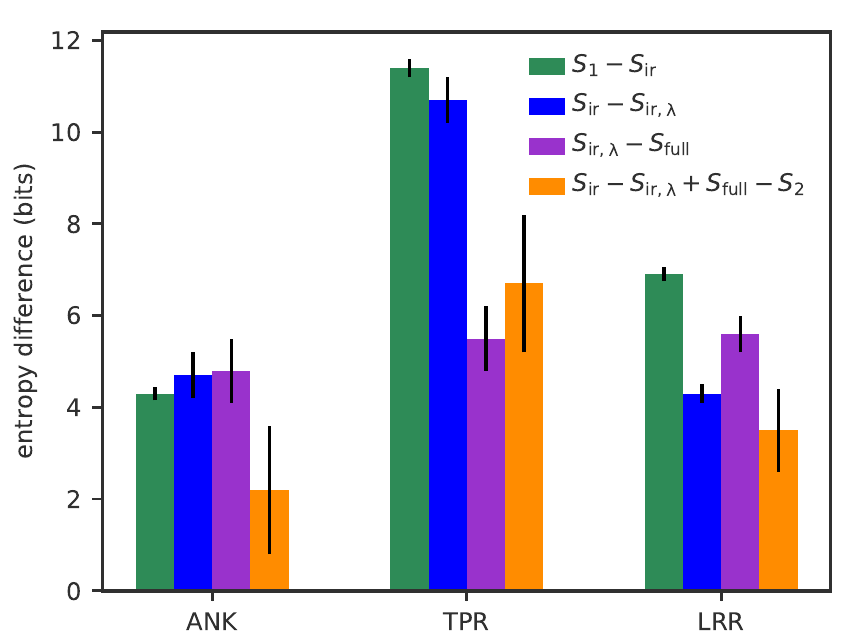}
\caption{
Contributions of within-repeat interactions ($S_1-S_{\rm ir}$ green), repeat-repeat interactions ($S_{\rm ir,\lambda}-S_{\rm full}$, purple), and phylogenic bias between consecutive repeats ($S_{\rm ir}-S_{\rm ir,\lambda}$, blue), to the entropy reduction from an independent-site model. All three contributions are comparable, but with a larger effect of within-repeat interactions and phylogenic bias in TPR. The fourth bar (orange) quantifies the redundancy between two constraints with overlapping scopes:
 the constraint on consecutive-repeat similary, and the constraint on repeat-repeat correlations. This redundancy is naturally measured within information theory by the difference of impact (i.e. entropy reduction) of a constraint depending on whether or not the other constraint is already enforced.
\label{thefig}
}
\end{figure}

The resulting entropies and their differences are reported in Table~\ref{table:entropies} and Fig.~\ref{thefig}.
All three considered families (ankyrins (ANK), leucine-rich repeats (LRR), and tetratricopeptides (TPR)) show a large reduction in entropy ($\sim 40-50\%$) compared to random polypeptide string models  of the same length $2L$ (of entropy $S_{\rm rand}=  2L\ln(21)$). Interactions and phylogenic similarity between repeats generally have a noticeable effect on family diversity, although the magnitude of this effect depends on the family: $(S_1-S_{\rm full})/S_{\rm full}=7\%$ for ANK, versus,  $13\%$ for LRR, and $16\%$ for TPR. Thus, although interactions are essential in correctly predicting the folding properties, they seem to only have a modest effect on constraining the space of accessible proteins compared to that of single amino-acid frequencies. However, when converted to numbers of sequences, this reduction is substantial, from $e^{\rm S_1}\sim 3\cdot 10^{54}$ to $e^{S_{\rm full}}\sim 2\cdot 10^{50}$ for ANK, from $10^{39}$ to $10^{34}$ for LRR, and from $7\cdot 10^{50}$ to $4\cdot 10^{42}$ for TPR. 

By considering models with more and more constraints, and thus with lower and lower entropy, we can examine more finely the contribution of each type of correlation to the entropy reduction, going from $E_1$ to $E_{\rm ir}$ to $E_{\rm ir,\lambda}$ to $E_{\rm full}$.
This division allows us to quantify the relative importance of phylogenic similarity between consecutive repeats ($\lambda_{\rm ID}$) relative to the impact of functional interactions ($J_{ij}$), as well as the relative weights of repeat-repeat versus within-repeat interactions (Fig.~\ref{thefig}).
We find that phylogenic similarity contributes substantially to the entropy reduction, as measured by $S_{\rm ir}-S_{\rm ir,\lambda}=4.5$ bits for ANK, 4.3 bits for LRR, and 10.7 bits for TPR. 
The contribution of repeat-repeat interactions ($S_{\rm ir,\lambda}-S_{\rm full}\sim 5$ bits for all three families) is comparable or of the same order of magnitude as that of within-repeat interactions ($S_1-S_{\rm ir}=4.3$ bits for ANK, 6.9 bits for LRR, and 11.4 bits for TPR). This result emphasizes the importance of physical interactions between neighboring repeats in the whole protein.

On a technical note, we also find that pairwise interactions encode constraints that are largely redundant with the constraint of phylogenic similarity between consecutive repeats, as can be measured by the double difference $S_{\rm ir}-S_{\rm ir,\lambda}-S_2+S_{\rm full}>0$ (Fig.~\ref{thefig}, orange bars). \rev{This redundancy comes from the fact that, in absence of an explicit constaint on $P({\rm ID})$ in $E_2$, the interaction couplings $J_{i,i+L}(\sigma,\sigma)$ between homologous positions in the two repeats is expected to favor pairs of identical residues to mimic the effect of $\lambda_{\rm ID}$.}
This redundancy motivates the need to correct for this phylogenic bias before estimating repeat-repeat interactions.

Comparing the three families, ANK has little phylogenic bias between consecutive repeats, and relatively weak interactions. 
By contrast, TPR has a strong phylogenic bias and strong within-repeat interactions.

\ssection{Effect of interaction range}

\begin{figure}
\includegraphics[width=\linewidth]{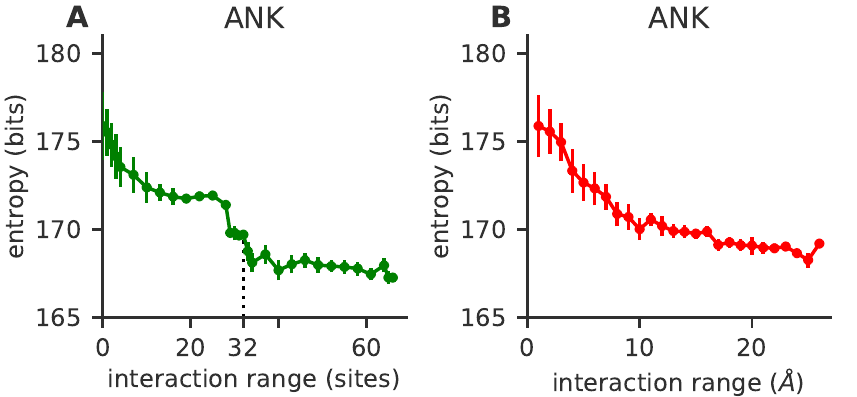}
\caption{Entropy reduction as a function of the range of interactions between residue sites.
A) Entropy of two consecutive ANK repeats, as a function of the maximum allowed interaction distance $W$ along the linear sequence. The entropy of the model decreases as more interactions are added and they constrain the space of possible sequences. After a sharp initial decrease at short ranges, the entropy plateaus until interactions between complementary sites in neighbouring repeats lead to a secondary sharp decrease at $W=L - 1=32$ (dashed line), due to structural interactions between consecutive repeats.
B) Entropy of two consecutive ANK repeats as a function of the maximum allowed three-dimensional interaction range. The entropy decreases rapidly until $\sim 10$ Angstrom, after which decay becomes slower. In both panels entropies are averaged over 10 realizations of fitting the model; see section~\ref{J_curves} and  for details of the learning and entropy estimation procedure. Error bars are estimated from fitting errors between the data and the model; see \mref{entropy_error} and Fig.~\ref{fig3_errbarsreals:fig} for error bars calculated as standard deviations over 10 realizations of model fitting.
 \label{theotherfig:fig}
}
\end{figure}

We wondered whether interactions constraining the space of accessible proteins had a characteristic lengthscale. To answer this question, for each protein family in Fig.~\ref{Fig0}, we learn a sequence of models of the form Eq.~\ref{H_ro}, in which $J_{ij}$ was allowed to be non-zero only within a certain interaction range $d(i,j)\leq W$, where  the distance $d(i,j)$ between sites $i$ and $j$ can be defined in two different ways: either the linear distance $|i-j|$ expressed in number of amino-acid sites, or the three-dimensional distance 
\rev{
between the closest heavy atoms
}
in the reference
structure of the residues.
Details about the learning procedure and error estimation are given in the Methods; see also Fig.~\ref{fig3_errbarsreals:fig} for an alternative error estimate.

The entropy of all families decreases with interaction range $W$, both in linear and three-dimensional distance, as more constraints are added to reduce diversity (Fig.~\ref{theotherfig:fig} for ANK, and Fig.~\ref{cutoff_LRR_TPR:fig} for LRR and TPR). The initial drop as a function of linear distance (Fig.~\ref{theotherfig:fig}A) is explained by the many local interactions between nearby residues in the sequence. The entropy then plateaus until interactions between same-position residues in consecutive repeats are included in the $W$ range, which leads to a sharp entropy drop at $W=L$. This suggests that long range interactions along the sequence generally do not constrain the protein ensemble diversity, except for interactions at exactly the scale of the repeat. This result suggests that the repeat structure is an important constraint limiting protein sequence exploration. These observations hold for all three repeat protein families. 
\rev{
The importance of 3D structure in reducing the entropy can also be appreciated in the entropy decay as a function of physical distance  (Fig.~\ref{theotherfig:fig}B for ANK) where most of the entropy drop happens within the first 10 angstroms, indicating that above this  characteristic distance 
interactions are not crucial in constraining the space of accessible sequences.
}

 % diff_families
\ssection{Multi-basin structure of the energy landscape}

\begin{figure}
\includegraphics[width=1.\linewidth]{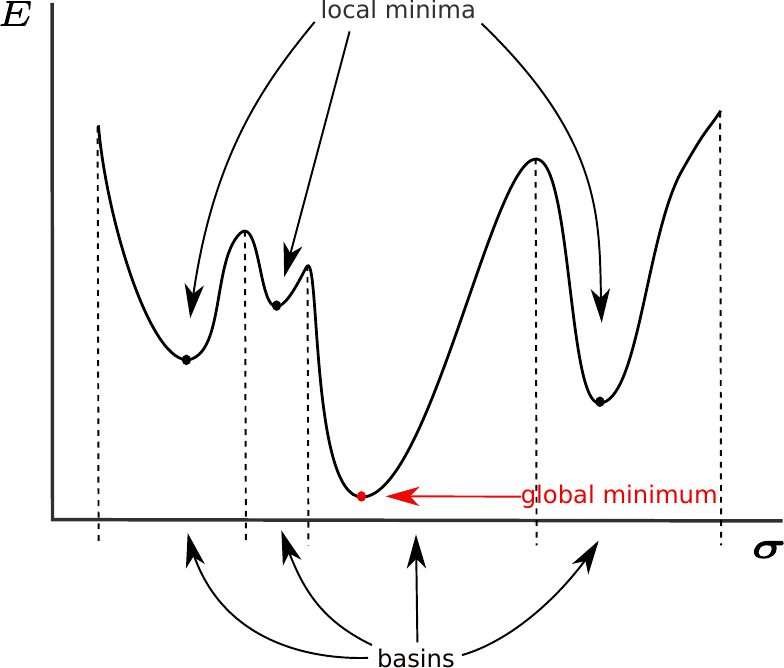}
\caption{ A rugged energy landscape is characterized by the presence of local minima, where proteins sequences can get stuck during the evolutionary process. The set of sequences that  evolve to a given local minimum defines the {\em basin of attraction} of that mimimum.
\label{rugged:fig}
}
\end{figure}

The energy function of Eq.~\eqref{H_ro} takes the same mathematical form as a disordered Potts model. These models, in particular in cases where $\sigma_i$ can only take two values, have been extensively studied in the context of spin glasses \cite{Mezard}. In these systems, the interaction terms $-J_{ij}(\sigma_i,\sigma_j)$ imply contradictory energy requirements, meaning that not all of these terms can be minimized at the same time --- a phenomenon called frustration. Because of frustration, natural dynamics aimed at minimizing the energy are expected to get stuck into local, non-global energy minima (Fig.~\ref{rugged:fig}), significantly slowing down thermalization. This phenomenon is similar to what happens in structural glasses in physics, where the energy landscape is ``rugged'' with many local minima that hinder the dynamics. Incidentally, concepts from glasses and spin glasses have been very important for understanding protein folding dynamics~\cite{Bryngelson7524}.

We asked whether the energy landscape of Eq.~\eqref{H_ro} was rugged with multiple minima, and investigated its structure. To find local minima, we performed a local energy minimization of $E_{\rm full}$ (learned with all constraints including on $P(\rm ID)$, but taken with $\lambda_{\rm ID}=0$ to focus on functional energy terms). By analogy with glasses, such a minimization is sometimes called a zero-temperature Monte-Carlo simulation or  a ``quench''.
The minimization procedure was started from many initial conditions corresponding to naturally occuring sequences of consecutive repeat pairs. At each step of the minimization, a random beneficial (energy decreasing) single mutation is picked; double mutations are allowed if they correspond to twice the same single mutation on each of the two repeats. Minimization stops when there are no more beneficial mutations. This stopping condition defines a local energy minimum, for which any mutation increases the energy. The set of sequences which, when chosen as initial conditions, lead to a given local minimum defines the {\em basin of attraction} of that energy mimimum (Fig.~\ref{rugged:fig}). The size of a basin corresponds to the number of natural proteins belonging to that basin.

\begin{figure}
\includegraphics[width=1.\linewidth]{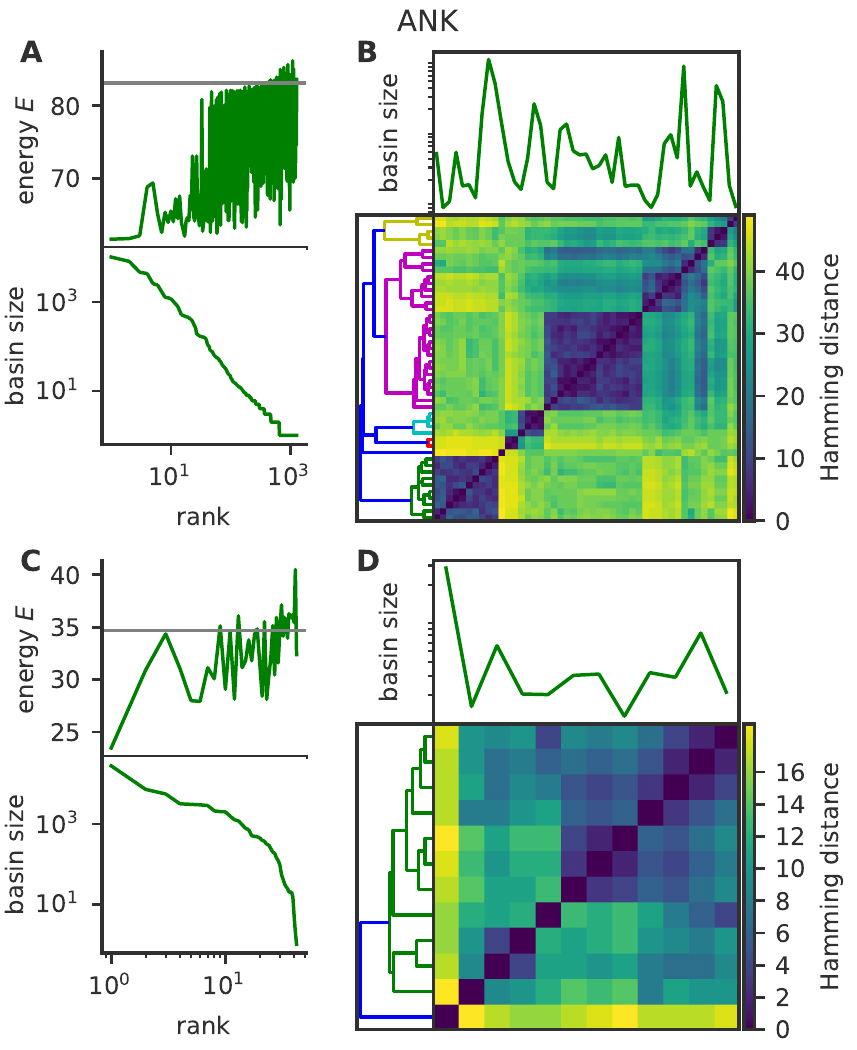}
\caption{Interactions within and between repeats sculpt a rugged energy landscape with many local minima. Local minima were obtained by performing a zero-temperature Monte-Carlo simulation with the energy function in Eq.~\eqref{H_ro}, starting from initial conditions corresponding to naturally occurring sequences of pairs of consecutive ANK repeats. A, bottom) Rank-frequency plot of basin sizes, where basins are defined by the set of sequences falling into a particular minimum. A, top) energy of local minima {\em vs} the size-rank of their basin, showing that larger basins often also have lowest energy. Gray line indicates the energy of the consensus sequence, for comparison.
  B) Pairwise distance between the minima with the largest basins (comprising 90\% of natural sequences), organised by hierarchical clustering. \rev{ The panel right above the matrix shows the the size of the basins relative to the minima corresponding to the entries of the distance matrix. }  A clear block structure emerges, separating different groups of basins with distinct sequences.
  C-D) Same as A) and B) but for single repeats.Since single repeats are shorter than pairs (length $L$ instead of $2L$), they have fewer local energy minima, yet still show a rich multi-basin structure. Equivalent analyses for LRR and TPR are shown in  Figs.~\ref{fig4_lrr:fig} and \ref{fig4_tpr:fig}.
\label{fig4:fig}
}
\end{figure}

Performing this procedure on \rev{natural sequences of} consecutive repeat  from all three families yielded a large number of local minima (Fig.~\ref{fig4:fig}). \rev{To control for the phylogenetic bias that links natural sequences, we repeated this analysis on sequences synthetically generated from the model ($E_{\rm full}$), and obtained very similar (see Fig.~\ref{fig4_synth:fig} for ANK)}. When ranked from largest to smallest, the distribution of basin sizes follows a power law (Fig.~\ref{fig4:fig}A for ANK and Fig.~\ref{fig4_lrr:fig}A and Fig.~\ref{fig4_tpr:fig}A for LRR and TPR). The energy of the minimum of each basin generally increases with the rank, meaning that largest basins are also often the lowest. Despite this multiplicity of local minima, the Monte-Carlo dynamics that we used in previous sections for learning the model parameters and for estimating the entropy did not get stuck in these minima, suggesting relatively low energy barriers between them.

The partition of sequences into basins allows for the definition of a new kind of entropy $S_{\rm conf}=-\sum_b P(b)\ln P(b)$ called configurational entropy, based on the distribution of basin sizes, $P(b)=\sum_{\bs\in b}P(\bs)$, where $\bs\in b$ means that energy minimization starting with sequence $\bs$ leads to basin $b$. This configurational entropy measures the effective diversity of basins, and is thus much lower than the sequence entropy $S_{\rm full}$, while the difference $S_{\rm full}-S_{\rm conf}$ measures the average diversity of sequences within each basin. We find $S_{\rm conf}=$5.1 bits for ANK, 6.0 bits for LRR, and 10.4 bits for TPR. As each basin corresponds to a distinct sub-family within each family \cite{Dokholyan2001}, this entropy quantifies the effective number of these subgroups.

While basins are very numerous, they are also not independent of each other. An analysis of pairwise distances (measured as the Hamming distance between the local minima) between the largest basins reveals that they can be organised into clusters (panels B of Figs.~\ref{fig4:fig}, \ref{fig4_lrr:fig}, and \ref{fig4_tpr:fig}), suggesting a hierarchical structure of basins, as is common in spin glasses \cite{Mezard}.

The impact of repeat-repeat interactions on the multi-basin structure can be assessed by repeating the analysis on the model of non-interacting repeats, $E_{\rm ir}$. In that model the two repeats are independent, so it suffices to study local energy minima of single repeats --- local minima of pairs of repeats follow simply from the combinatorial pairing of local minima in each repeat. The analyses of basin size distributions, energy minima, and pairwise distances in single repeats are shown in panels C and D of Figs.~\ref{fig4:fig}, \ref{fig4_lrr:fig}, and \ref{fig4_tpr:fig}. We still find a substantial number of unrelated energy minima, suggesting again several distinct subfamilies even at the single-repeat level. For comparison, the configurational entropy of pairs of independent repeats is 6.9 bits for ANK, 6.7 for LRR, and 7.6 for TPR. While for ANK and LRR repeat-repeat interactions decrease the configurational entropy, as they do for the conventional entropy, they in fact increase entropy for TPR, making the energy landscape even more frustrated and rugged.

\rev{
Note that the independent sites model $E_1$ defines a convex energy landscape with a single local minimum --- the consensus sequence --- as all constraints $h_i$ can be optimized independently.
To address how  the interactions contribute in shaping the sequence space, going from a convex to a rugged landscape, we repeated the analysis with a limited linear interaction range $W$ of 3 and 10 (models of Fig.~\ref{theotherfig:fig} A). We find that the more interactions we add, the more local minima we find (Fig.~\ref{fig4_W:fig}A and B for ANK with $W=3$, and C and D for $W = 10$). The minima cluster into clearer sub-blocks structure as the interaction range is increased, consistent with the entropy reduction in entropy observed in~\ref{theotherfig:fig} A.}

In summary, the analysis of the energy landscape reveals a rich structure, with many local minima ranging many different scales, and with a hierarchical structure between them.

\ssection{Distance between repeat families}

Lastly, we compared the statistical energy landscapes of different repeat families. Specifically, we calculated the Kullback-Leibler divergence between the probability distributions $P(\bs)$ (given by Eqs.~\ref{eq:boltzmann}-\ref{H_ro}) of two different families, after aligning them together in a single multiple sequence alignment (see \mref{methods:DKL}).

We find essentially no similarity between ANK and TPR, despite them having similar lengths: $D_{\rm KL}({\rm ANK} || {\rm TPR}) = 227.6$ bits, and $D_{\rm KL}({\rm TPR} || {\rm ANK}) = 214.1$ bits. These values are larger than the Kullback-Leibler divergence between the full models for these families and a random polypetide, $D_{\rm KL}({\rm ANK} || {\rm rand}) =122.8$ bits, and $D_{\rm KL}({\rm TPR} || {\rm rand}) =157.6$ bits. LRR is not comparable to ANK or TPR as it is much shorter, and a common alignment is impractical. These large divergences between families of repeat proteins
show that different families impose quantifiably different constraints, which have forced them to diverge into different troughs of non-overlapping energy landscapes. This lack of overlap makes it impossible to find intermediates between the two families that could evolve into proteins belonging to both families.

\section{Discussion}

 % discussion

Our analysis of repeat protein families shows that the constraints between amino acids in the sequences allows for an estimation of the size of the accessible sequence space.  The obtained numbers (ranging from 141 bits to 167 bits, corresponding to $10^{36}$ to $10^{50}$ sequences) are of course huge compared to the number of sequences in our initial samples ($\sim20,500$ for ANK,  $\sim 18,800$ for LRR, and $\sim 10,000$ for TPR), but comparable to the total number of proteins having been explored over the whole span of evolution, estimated to be $10^{43}$ in Ref.~\cite{Dryden2008}.

In particular, we have quantified the reduction of the accessible sequence space with respect to random polypeptides.
While most of this reduction is attributable to conservation of residues at each site,
interactions between amino acids, both within and between consecutive repeats, significantly constrain the diversity of all repeat families. The break-up of entropy reduction between the three different sources of constraints --- within-repeat interactions, between-repeat interactions, and evolutionary conservation between consecutive repeats --- is fairly balanced, although TPR stands out as having more within-repeat interactions and more conservation between neighbours, suggesting that it may have had less time to equilibrate. 

All studied repeat families have rugged energy landscapes with multiple local energy minima. Note that the emergence of this multi-valley landscape is a consequence of the interactions between amino acids: models of independent positions ($E_1$) only admit a single energy minimum corresponding to the consensus sequence.
This multiplicity of minima allow us to collapse multiple sequences to a small number of coarse-grained attractor basins. These basins suggest that mutations between sequences within one coarse-grained basin are much more likely than mutating into sequences in other basins. In general, our results paint a picture of further subdivisions within a family, and define sub-families due to the fine grained interaction structure. \rev{Going beyond single families, this analysis suggest a view in which natural proteins all live in a global evolutionary landscape, of which families would be basins, or clusters of basins, with a hierarchical  structure \cite{Dokholyan2001}.}

This overall picture of the sequence energy landscape is reminiscent of the hierarchical picture of the structural energy landscape of globular proteins, an overall funneled shape with tiers within tiers ~\cite{pmid1749933}.  
The form of the energy landscape forcibly shapes the accessible evolutionary paths between sequences. The rugged and further subdivided structure shows that the uncovered constraints are global, and not just pairwise between specific residues. Therefore even changing two residues together, as is often done in laboratory experiments, is not enough to recover the evolutionary trajectories. \rev{While other approaches have explored local accessible directions of evolution \cite{Facco2019}, our results suggest more global, non local modes of evolution between clusters.}

Interestingly, the sequences that correspond to the energy minima of the landscapes are not found in the natural dataset. This observation can be either due to sampling bias (we have not yet observed the sequence with the minimal energy, although it exists), or this sequence may not have been sampled by nature.
Alternatively, there may be additional functional constraint that are not included in our model to avoid these low energy sequences (e.g. a too stable protein may be difficult to degrade).

Even more intriguingly, sequences with minimal energy do not correspond to the consensus sequence of the alignment (whose energy is marked by a gray line in panel A of Figs.~\ref{fig4:fig}, \ref{fig4_lrr:fig}, and \ref{fig4_tpr:fig}), suggesting that the consensus sequence can be improved upon.
All three repeat protein families studied here have been shown to be amenable to simple consensus-guided design of synthetic proteins. Synthetic proteins based on the consensus sequences of multiple alignments~\cite{pmid21715155}  were found to be foldable and very stable against chemical and thermal denaturation. Mutations towards consensus amino acids in the ANK family members have been experimentally shown to both stabilize the whole repeat-array and they may tune the folding paths towards nucleating folding in the consensus sites~\cite{pmid18396879, Barrick2008} . Our results suggest that interactions may play an additional role in stabilizing the sequences, and propose alternative solutions to the consensus sequences in the design of synthetic proteins.

\section{Methods}

 % methods_jacopo
\beginsupplement

\newcommand{\captionSone}{
{\bf Reproducibility of entropy estimation.}
Entropy as a function of the maximum linear interaction range $W$ along the sequence.  Green curve: entropy of the ANK family with error bars calculated as standard deviations over 10 model learning realizations, where models are learned by incrementally adding more interaction terms as $W$ is increased, taking the model learned at $W-1$ as initial condition.
This plot is the same as in Fig.~\ref{theotherfig:fig}A but with the different error bar estimates, showing that our results are robust to the details of error estimation. 
Red curve: entropy obtained after {\em de novo} learning for each $W$, starting from a non-interacting model as initial condition.
With those initial conditions the learning gets stuck, leading to systematically overestimating the entropy and missing the second entropy drop at $W=L - 1$. See section~\ref{J_curves} for details of the learning and entropy estimation procedure.
}

\newcommand{\captionStwo}{
{\bf Range dependence of entropy in LRR and TPR families.}
Entropy of the LRR (A) and TPR (B) family as a function of the maximum interaction distance $W$ along the sequence.   The entropy of the model decreases as a more interactions are added and they constrain the space of possible sequences. As with ANK, the entropy first drops, plateaus, then drops again at the distance corresponding to homologous positions along the two repeats ($W=L - 1=23$ for LRR, and $33$ for TPR, dashed line). This second drop indicates that there is a typical  distance along the sequence, corresponding to the repeat length, where  interactions due to  structural properties  constrain  the sequence ensemble.  The error bars are estimated approximately from errors in learning (see Section~\ref{entropy_error}).
Entropies are averaged over 5 realizations of the learning and entropy estimation procedure.
}

\newcommand{\captionSthree}{
{\bf Analysis of local energy minima for pairs of consecutive repeats of LRR.} Energy minima were obtained by zero-temperature dynamics. Sequences falling into a given minimum with these dynamics define its basin of attraction.
A, bottom) rank-frequency plot of the sizes of the basins of attraction. A, top) energy minimum of each basin. Gray line shows the energy of the consensus sequence B) Pairwise Hamming distances between energy minima, organised by hierarchical clustering. \rev{ The panel right above the matrix shows the the size of the basins relative to the minima corresponding to the entries of the distance matrix. }  
C and D) Same analysis as A) and B), but for single LRR repeats.
}

\newcommand{\captionSfour}{
{\bf Analysis of local energy minima for pairs of consecutive repeats of TPR.} Energy minima were obtained by zero-temperature dynamics. Sequences falling into a given minimum with these dynamics define its basin of attraction.
A, bottom) rank-frequency plot of the sizes of the basins of attraction. A, top) energy minimum of each basin. Gray line shows the energy of the consensus sequence B) Pairwise Hamming distances between energy minima, organised by hierarchical clustering. \rev{ The panel right above the matrix shows the the size of the basins relative to the minima corresponding to the entries of the distance matrix. }  
C and D) Same analysis as A) and B), but for single TPR repeats.
}

\newcommand{\captionSfive}{
\rev{{\bf Interactions within repeats increase the ruggedness of the energy landscape.} Local minima were obtained by performing a zero-temperature Monte-Carlo simulation with the energy function in Eq.~\eqref{H_ro} with non-zero $J_{ij}$ within  linear interaction range $W$, starting from initial conditions corresponding to naturally occurring sequences of pairs of consecutive ANK repeats, for $W=3$ (A and B) and $W=10$ (C and D). See Fig.~\ref{fig4:fig} for the full model ($W=2L$).
A and C, bottom: Rank-frequency plot of basin sizes, where basins are defined by the set of sequences falling into a particular minimum. A and C, top: energy of local minima {\em vs} the size-rank of their basin. Gray line indicates the energy of the consensus sequence, for comparison.
  B and D: Pairwise distance between the minima with the largest basins (comprising 90\% of natural sequences), organised by hierarchical clustering. The panel right above the matrix shows the the size of the basins relative to the minima corresponding to the entries of the distance matrix. The  block structure starts emerging as interactions are turned on (D versus B).}
}

\newcommand{\captionSsix}{
\rev{{\bf Analysis of local energy minima from generated pairs of consecutive repeats of ANK}. Energy minima were obtained by zero-temperature dynamics starting from sequences generated {\em in silico} from $E_{\rm full}$. Sequences falling into a given minimum with these dynamics define its basin of attraction.
A, bottom) rank-frequency plot of the sizes of the basins of attraction. A, top) energy minimum of each basin. Gray line shows the energy of the consensus sequence B) Pairwise Hamming distances between energy minima, organised by hierarchical clustering.  The panel right above the matrix shows the the size of the basins relative to the minima corresponding to the entries of the distance matrix.   
C and D) Same analysis as A) and B), but for single ANK repeats.}
}

\ssection{Data curation}~\label{data}

We use a previously curated alignment of pairs of repeats for each family ~\cite{Espada2018}: ANK (PFAM id PF00023 with a final alignment of 20513 sequences of $L = 66$ residues each), LRR (PFAM id PF13516  with a final alignment of 18839 sequences of $L = 48$ residues each) and TPR (PFAM id PF00515  with a final alignment of 10020 sequences of $L = 68$ residues each). Those multiple sequence alignments of repeats were obtained from PFAM 27.0~\cite{bateman2004pfam,finn2013pfam}. In order to improve the data obtained from the PFAM database, we used original full protein sequences available in UniProt database~\cite{TheUniProtConsortium01012014} to  
add available information using the headers of the original alignement. Firstly, to decrease the number of gaps positions, misdetected initial and final amino acids in repeats were completed with residues from full sequences. Secondly, individual repeats which appeared consecutively in natural proteins were joined into pairs. Finally, positions with more than 80\% of gaps along the alignment were removed, eliminating in this way insertions. 

From the multiple sequence alignement of each family, they were calculated the observables that we use to constrain our statistical model. Particularly, we calculated the marginal frequency $f_i(\sigma_i)$ of an amino acid $\sigma_i$ at position $i$ and the joint frequency $f_{ij}(\sigma_i, \sigma_j)$ of two amino acids $\sigma_i$ and $\sigma_j$ at two different positions $i$ and $j$. These quantities were calculated using only sequences selected by clustering at $90\%$ of identity computed with CD-HIT~\cite{cdhit2002} and then normalizing by the amount of sequences. In this way, the occurrences of residues in every position are not biased by overrepresentation of  proteins in the database. Furthermore, to take into account the repeated nature of the protein families that we are considering, an additional observable was calculated, the distribution of sequence overlap between two consecutive repeats, $P({\rm ID}(\bs))$, with ${\rm ID} (\bs)=\sum_{i=1}^L \delta_{\sigma_i,\sigma_{i+L}}$. 

\ssection{Model fitting}~\label{model_fitting}

In order to obtain a model that reproduces the experimentally observed site-dependent amino-acid  frequencies, $f_i(\sigma_i)$, correlations between two positions, $f_{ij}(\sigma_i, \sigma_j)$, and the distribution of Hamming distances between consecutive repeats, $P({\rm ID}(\bs))$, we apply a likelihood gradient ascent procedure, starting from an initial guess of the $ h_i(\sigma_i)$, $J_{ij}(\sigma_i, \sigma_j)$ and $\lambda_{\rm ID}(\bs)$ parameters.

At each step, we  generate 80000 sequences of length $2L$ through a Metropolis-Hastings Monte-Carlo sampling procedure. We start from a random amino-acid sequence and we  produce many point mutations in any position, one at a time. If a mutation decreases the energy \eqref{H_ro} we accept it. If not, we accept the mutation with probability $e^{-\Delta E}$, where $\Delta E$ is the difference  of energy between the original and the mutated sequence.  We add one sequence to our final ensemble every $1000$ steps. 
Once we generated the sequence ensemble, we measure its marginals $f^{\rm model}_i(\sigma_i)$ and $f^{\rm model}_{ij}(\sigma_i, \sigma_j)$, as well as $P^{\rm model}({\rm ID}(\bs))$, and update the parameters of  Eq.~\ref{H_ro} following the gradient of the likelihood.
The local field and $\lambda_{\rm ID}(\bs)$ are updated along the gradient of the per-sequence log-likelihood, equal to the difference between model and data averages:
\begin{equation}\label{h_upd} 
h_i(\sigma_i)^{t+1} \leftarrow h_i(\sigma_i)^{t} + \epsilon_m[f_i(\sigma_i) - f_i^{\rm model}(\sigma_i)],
\end{equation}
\begin{equation}\label{l_upd} 
\lambda_{\rm ID}(\bs)^{t+1} \leftarrow \lambda_{\rm ID}(\bs)^{t} - \epsilon_{\rm ID}[P(\rm ID(\bs)) - P(\rm ID(\bs))^{\rm model}].
\end{equation}
As the number of parameters for the interaction terms $J_{ij}$ is large ($= 21^2 L^2$), we force to 0 those that are not contributing significantly to the model frequencies through a $L_1$ regularisation $\gamma \sum_{ij,\sigma,\tau}|J_{ij}(\sigma,\tau)|$ added to the likelihood. This leads to the following rules of maximization:
If $J_{ij}(\sigma_i, \sigma_j)^{t} = 0 $ and $\abs{f_{ij}(\sigma_i, \sigma_j) - f_{ij}^{\rm model}(\sigma_i, \sigma_j)} < \gamma$
\begin{equation}\label{j_upd1} 
J_{ij}(\sigma_i, \sigma_j)^{t+1} \leftarrow 0.
\end{equation}
If $J_{ij}(\sigma_i, \sigma_j)^{t} = 0 $ and $\abs{f_{ij}(\sigma_i, \sigma_j) - f_{ij}^{\rm model}(\sigma_i, \sigma_j)} > \gamma$
\begin{equation}\label{j_upd2} 
\begin{split}
J_{ij}(\sigma_i, \sigma_j)^{t+1} \leftarrow   \epsilon_j[ & f_{ij}(\sigma_i, \sigma_j) - f_{ij}^{\rm model}(\sigma_i, \sigma_j) - \\
& \gamma {\rm sign}(f_{ij}(\sigma_i, \sigma_j) - f_{ij}^{\rm model}(\sigma_i, \sigma_j))].
\end{split}
\end{equation}
If $\Big[J_{ij}(\sigma_i, \sigma_j)^{t} + \epsilon_j[f_{ij}(\sigma_i, \sigma_j) - f_{ij}^{\rm model}(\sigma_i, \sigma_j) - \gamma {\rm sign}(J_{ij}(\sigma_i, \sigma_j)^{t})]\Big]  J_{ij}(\sigma_i, \sigma_j)^{t} \geq 0 $ 
\begin{equation}\label{j_upd3} 
\begin{split}
J_{ij}(\sigma_i, \sigma_j)^{t+1} \leftarrow   J_{ij}(\sigma_i, \sigma_j)^{t} + \epsilon_j[& f_{ij}(\sigma_i, \sigma_j) - f_{ij}^{\rm model}(\sigma_i, \sigma_j) - \\
& \gamma {\rm sign}(J_{ij}(\sigma_i, \sigma_j)^{t})].
\end{split}
\end{equation}
If $\Big[J_{ij}(\sigma_i, \sigma_j)^{t} + \epsilon_j[f_{ij}(\sigma_i, \sigma_j) - f_{ij}^{\rm model}(\sigma_i, \sigma_j) - \gamma {\rm sign}(J_{ij}(\sigma_i, \sigma_j)^{t})]\Big]  J_{ij}(\sigma_i, \sigma_j)^{t} < 0 $ 
\begin{equation}\label{j_upd4} 
J_{ij}(\sigma_i, \sigma_j)^{t+1} \leftarrow  0.
\end{equation}

The optimization parameters were set to: $\epsilon_m=0.1$, $\epsilon_j=0.05$, $\epsilon_{\rm ID}=10$, and $\gamma=0.001$. 

To estimate the model error, we compute  $f_i(\sigma_i) - f_i^{\rm model}(\sigma_i)$ and $f_{ij}(\sigma_i, \sigma_j) - f_{ij}^{\rm model}(\sigma_i, \sigma_j)$. We also calculate the difference of generated and natural repeat similarity distribution for all the possible repeats Hamming distances, penalized by a factor 5 to better learn the parameter $\lambda_{\rm ID}$:  $5(P(\rm ID(\bs)) - P(\rm ID(\bs))^{\rm model})$. 
We repeat the procedure above until the maximum of all errors, $|f_i(\sigma_i) - f_i^{\rm model}(\sigma_i)|$, $|f_{ij}(\sigma_i, \sigma_j) - f_{ij}^{\rm model}(\sigma_i, \sigma_j)|$ and $5|P(\rm ID(\bs)) - P(\rm ID(\bs))^{\rm model}|$, goes below $0.02$,  as in Ref.~\cite{Espada2018}.

\ssection{Models with different sets of constraints}\label{J_curves}

Using this procedure we can calculate the model defined in Eq.~\ref{H_ro} with different interaction ranges used in the entropy estimation in Fig.~\ref{theotherfig:fig} A. We  start from the independent model $h_i(\sigma_i)= \log{f_i(\sigma_i)}$. We first learn the model in  Eq.~\ref{H_ro}  with $J=0$. 
We then re-learn models with interactions between sites $i,j$ along the linear sequence such that $\abs{i -j}\leq W$, 
in a seeded way starting from the previous model.
The first and last point of Fig.~\ref{theotherfig:fig} correspond to the independent site model with $\lambda_{\rm ID}$ and the full model in Eq.~\ref{H_ro}

The entropy in Fig.~\ref{theotherfig:fig}B is calculated in the same way as in Fig.~\ref{theotherfig:fig}, but now interactions are turned on progressively according to physical distance in the 3D structure rather than the linear sequence distance. 
In order to obtain the physical distance between residues we use as a reference structure the first two repeats of a consensus designed ankyrin protein 1n0r~\cite{peng2002pnas, pluckthun2003jmb}, which have exactly $66$ amino-acids. We define the 3D separation between two residues 
\rev{
as the minimum distance between their heavy atoms
}
in the reference structure.

To learn the Potts model without $\lambda_{\rm ID}$ ($E_{2}$)  
we  remove $\lambda_{\rm ID}$ from Eq.~\ref{H_ro} and re-learn the Potts field using the full model parameters as initial contition. 
  
To learn the single repeat models with and without $\lambda$ ($E_{\rm ir}$ and $E_{\rm ir,\lambda}$, we take as initial condition the model with interactions below the length of a repeat ($W = L - 1$, dashed vertical line in Fig.~\ref{theotherfig:fig}), and then learn a model removing all the $J_{ij}$ terms between different repeats. We also impose that the $h_i$  fields and intra-repeats $J_{ij}$ terms are the same in each repeat, and the experimental amino-acid frequencies to be reproduced by the model are the average over the two repeats of the 1- and 2-points intra-repeats frequencies $f_i(\sigma_i)$ and $f_{ij}(\sigma_i, \sigma_j) $, such that 
  \beq
 f'_i(\sigma_i)=f'_{i+L}(\sigma_i) = \frac{1}{2}\Big(f_i(\sigma_i)+f_{i+L}(\sigma_i)\Big),
 \eeq 
 and 
 \begin{eqnarray}
 f'_{ij}(\sigma_i, \sigma_j)&=& f'_{i+L, j+L}(\sigma_i, \sigma_j) =\\ \nonumber
 &=& \frac{1}{2} (f_{ij}(\sigma_i, \sigma_j)+f_{i+L, j+L} (\sigma_i, \sigma_j) ),
 \end{eqnarray}
 if $i$ and $j$ represent sites within the same repeat. In this way we obtain a model for a single repeat that can be extended to both the repeats in the original set of sequences of our dataset.

\ssection{Entropy estimation}\label{entropy_calc_meth:sec}

In practice to calculate the entropy $S$ of the protein families we relate it to the  internal energy $E= - \log{p(\bs)}$ and the free energy $F= - \log{Z}$:
\begin{align}\label{S_can} 
S &= \langle E \rangle - F \nonumber  \\
&= \sum_{\bs} p(\bs) E(p(\bs)) + \log{Z} \\
 &= - \sum_{\bs} p(\bs) \log{p(\bs)} \nonumber \,,
\end{align}
We generate sequences according to the energy function  in  Eq.~\ref{H_ro} and use them to numerically compute $ \langle E \rangle$. To calculate the free energy we use the auxilliary energy function:
\begin{equation}\label{H_alpha} 
E_\alpha(\bs)= - \sum_i h_i(\sigma_i)+ \alpha \Big[ -\sum_{ij} J_{ij}(\sigma_i, \sigma_j)  + \lambda_{{\rm ID}} \Big] \,,
\end{equation}
where the interaction strength  across different sites can be tuned through a parameter $\alpha$ that is changed from $0$ to $1$. We generate protein sequence ensembles with different values of $\alpha$  and  use them to calculate $F$ as a function of $\alpha$, $F(1)= F(0) + \int_0^1{\rm d}\alpha \frac{{\rm d} F}{{\rm d}\alpha}$:
\begin{equation}\label{F_int} 
F(1)= F(0) + \int_0^1{\rm d}\alpha  \left\langle -\sum_{ij} J_{ij}(\sigma_i, \sigma_j)  + \lambda_{{\rm ID}}  \right\rangle_\alpha \,,
\end{equation}
where the average over $\alpha$ is taken over the sequences generated with a certain value of $\alpha$, characterized by the ensemble with probability $p_{\alpha}(\bs)=(1/Z_{\alpha})e^{-E_\alpha(\bs)}$.   $F(0)$ is the free energy for an independent sites model:
\begin{equation}\label{F_0} 
F(0)= - \sum_{i} \log{\sum_{\sigma_i} {\rm e}^{h_i(\sigma_i)}} \,,
\end{equation}
where the first sum is taken over protein sites and the second over all possible amino-acids at a given site. Eq.~\ref{F_0} and  Eq.~\ref{S_can} result in  the thermodynamic sampling approximation for calculating the entropy~\cite{Frankel2007}:
\begin{equation}\label{S_fin} 
S= \langle E \rangle + \sum_{i} \log{\sum_{\sigma_i} {\rm e}^{h_i(\sigma_i)}}  - \int_0^1{\rm d}\alpha \left\langle -\sum_{ij} J_{ij}(\sigma_i, \sigma_j)  + \lambda_{{\rm ID}}  \right\rangle_\alpha \,.
\end{equation}
We generate 80000 sequences using Monte Carlo sampling for the energy in Eq.~\ref{H_alpha} with 50 different $\alpha$ values, equally spaced between $0$ and $1$ at a distance of $0.02$, and then numerically compute the integral in Eq.~\ref{S_fin} using the Simpson rule. 

\ssection{Entropy error}\label{entropy_error}

The entropy estimate is subject to three sources of uncertainty: the finite-size of the dataset, convergence of parameter learning, and the noise in the thermodynamic integration. 
 We estimate the contribution of each of these errors using the independent sites model. In the independent sites model each site $i$ is simply described by a multinomial distribution with weights given by the observed amino-acid frequencies in the datasets. The variance in the estimation of the frequencies from a finite size sample is ${\rm Var}(f_i(\sigma_i) )= {\left(p_i(\sigma_i)(1 - p_i(\sigma_i))\right)}/{N_s}$ and the covariance between the frequencies of different amino-acids $\sigma$ and $\sigma'$ at the same site $i$ is ${\rm Cov}(f_i(\sigma_i), f_i(\sigma'_i) )=  -{\left(p_i(\sigma_i)p_i(\sigma'_i)\right)}/{N_s}$
where $N_s$ is the sample size and $p_i(\sigma_i)$ are the weights of the true multinomial distribution sampled.
Through error propagation from these quantities we calculate  the  variance in the entropy of the independent sites model, to first order in ${1}/{N_s}$:
\begin{equation}\label{S_var} 
\begin{split}
{\rm Var}(S_{indep})= & \frac{1}{N_s} \Big[ \sum_i \sum_{\sigma_i} p_i(\sigma_i) \log{p_i(\sigma_i)}^2 - S_{indep}^2 \Big] \\ 
&+ O(\frac{1}{N_s^2}) \,.
\end{split}
\end{equation}
Evaluating this equation using the empirical frequencies $p=f$
assuming they are sampled from an underlying multinomial distribution, gives an estimate of the standard deviation of $0.05$. We assume that the interaction terms do not change the order of magnitude of this estimation.
Also the standard deviation in the averages in Eq.~\eqref{S_fin} scales as ${1}/{\sqrt{N_s}}$ with $N_s=80000$. 

\begin{figure}
\includegraphics[width=\linewidth]{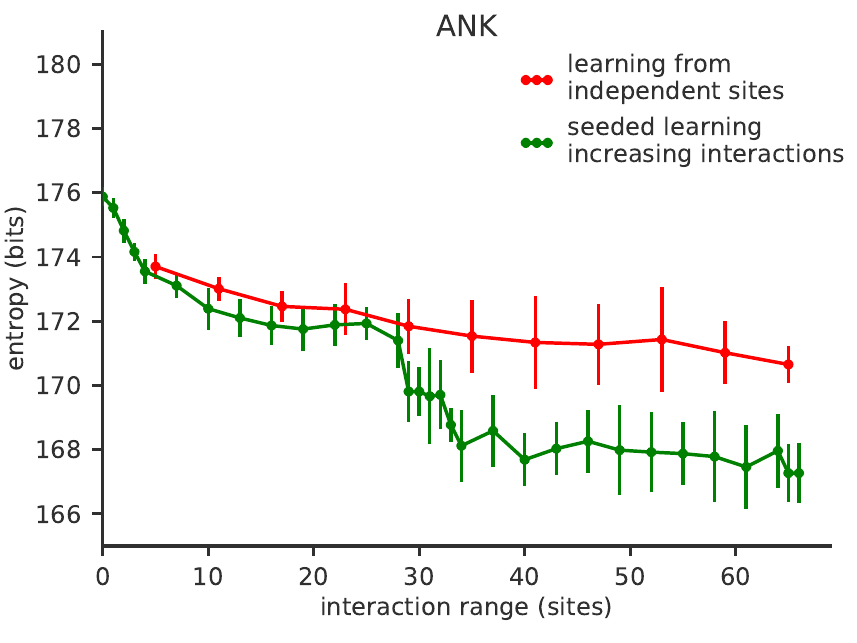}
\caption{\captionSone
 \label{fig3_errbarsreals:fig}
}
\end{figure}

The
parameter inference is affected not only by noise, but also by a systematic bias depending on the parameters of the gradient ascent described in Section~\ref{model_fitting} and the initial condition that we chose to start learning from. Fig.~\ref{fig3_errbarsreals:fig} shows the average entropy of 10 realizations of the learning and thermodynamic integration procedure for the ANK family and its standard deviation as error bars. If we learn the models with an increasing $W$ window progressively we get a different profile than learning each point starting from the independent model, and above $L$ these two profiles are more distant than the magnitude of the standard deviation, signalling  a systematic bias. Fig.~\ref{fig3_errbarsreals:fig} also shows that progressively  learning  the model results in a better  parameters convergence to values that  give lower  entropy values.

\begin{figure}
\includegraphics[width=\linewidth]{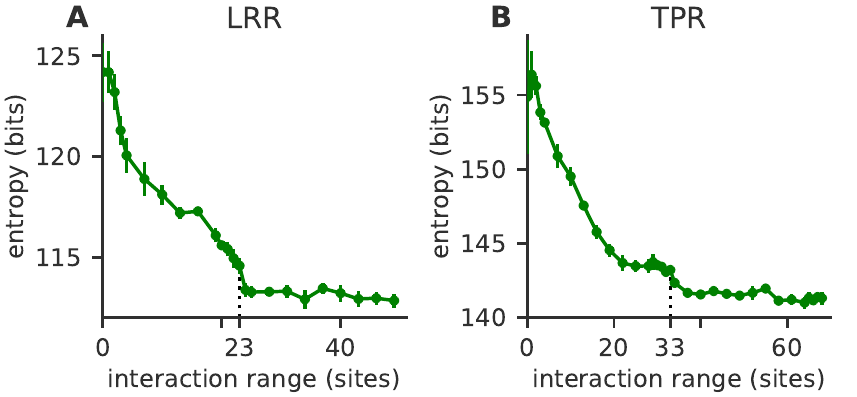}
\caption{
\captionStwo
\label{cutoff_LRR_TPR:fig}
}
\end{figure}

In order to estimate how this bias is reflected in the entropy estimation we take the single-site amino-acid frequencies produced by the inferred energy function in the last Monte-Carlo phase of the learning procedure 
 and calculate the corresponding entropy for this independent-sites model. We compute the absolute value of the difference between this estimate of the entropy and the independent-sites entropy calculated from the  dataset. Again in doing this we assume that neglecting the interaction terms does not change the order of magnitude of this error. 
 These procedure results in the errorbars shown in Fig.~\ref{theotherfig:fig},Fig.~\ref{thefig}, Table~\ref{table:entropies}, Fig.~\ref{cutoff_LRR_TPR:fig}. 

We repeat $10$ realizations of both the parameter inference procedure and the entropy estimation, and in Fig.~\ref{theotherfig:fig} we show the average entropy of these $10$ numerical experiments for the ANK family where error bars are estimated as explained above to sketch the order of magnitude of the error coming from systematic bias in the parameters learning. Fig.~\ref{fig3_errbarsreals:fig} shows the mean entropy of ANK as in Fig.~\ref{theotherfig:fig} A   with   the  standard deviations of the realizations entropy as error bars, to give an idea of the combined noise in the thermodynamic integration and in the gradient descent, starting from the same initial conditions and with the same update parameters (see Section~\ref{model_fitting}). The combined  noise is smaller than the entropy decrease at $33$ residues, showing the decrease is real. 

To further check the robustness of the entropy estimation procedure, we generate two synthetic ANK datasets, one with an independent sites model, the other with a model of two non-interacting repeats obtained as explained in the Section~\ref{model_fitting}, and relearn the model from the synthetic datasets. Repeating the learning and entropy estimation procedure on each on the synthetic protein families gives results that are consistent with the model used for the dataset generation. The entropy of the model learned taking an independent sites dataset does not decrease with the interaction range $W$  and the entropy of the model learned taking a non-interacting repeats dataset  does not show any drop around the repeat length.

We repeat the procedure described for the LRR and TPR repeat-proteins families such as LRR and TPR reaching similar conclusions (Fig.~\ref{cutoff_LRR_TPR:fig}). 

\ssection{Calculating the basins of attraction of the energy landscape}~\label{quench}

In order to characterize the ruggedness of the inferred energy landscapes and the sequence identity of the local minima,  we start from all the sequences in the natural dataset as initial conditions and for each of them we perform a $T=0$ quenched Monte-Carlo procedure. \rev{Repeating this analysis on sequences synthetically generated from $E_{\rm full}$ yields very similar results (see Fig.~\ref{fig4_synth:fig} for ANK)}

We perform this energy landscape exploration  learning the parameters of the Hamiltonian in~Eq.~\ref{H_ro} (refer to Section~\ref{model_fitting} for the learning procedure), and then set $\lambda_{\rm ID}=0$ in the energy function because we want to investigate the shape of the energy landscape due to selection rather than the phylogenic dependence. 

We scan all the possible mutations that decrease the sequence energy and then draw one of them from a uniform random distribution. 
The possible mutations are all single point mutations. If the same amino-acid is present in the same relative position in the two repeats we allow for double mutations that mutate those two positions to a new amino-acid, that is identical in both repeats, at the same time.  We do this so that the phylogenetic biases that are still partially present in the parameters of the model 
do not result in spurious local minima biasing the quenching results. The Monte-Carlo procedure ends when every proposed move results in a sequence with an increased energy, and the identified sequence is a local minimum of the energy landscape. 

\rev{
To explore how turning on interactions makes the energy landscape more rugged, we perform the same procedure with the Hamiltonian corresponding to two intermediate interaction ranges in  Fig.~\ref{theotherfig:fig} A. That is Eq.~\ref{H_ro}, in which $J_{ij}$ was allowed to be non-zero only within a certain interaction range $W$. We picked $W=3$ and $W=10$.
}

In order to assess what is the role of the inter-repeat interactions we repeat this $T=0$ quenched Monte-Carlo procedure on single repeats, with all the unique repeats in the natural dataset as initial condition. The learning procedure of the Hamiltonian for a single repeat is explained in Section~\ref{model_fitting}. In this single repeat case the  possible mutations are just the single point mutations.

Once we have the local minima of the energy landscape, we obtain the coarse-grained minima using the Python Scipy hierarchical clustering algorithm.
 In this hierarchical clustering the distance between two clusters is calculated as the average Hamming distance between all the possible pairs of sequences belonging each to one cluster. As a result we plot the clustered distance matrix, the clustering dendogram and the basin size corresponding to the distance matrix entries.

\begin{figure}
\includegraphics[width=\linewidth]{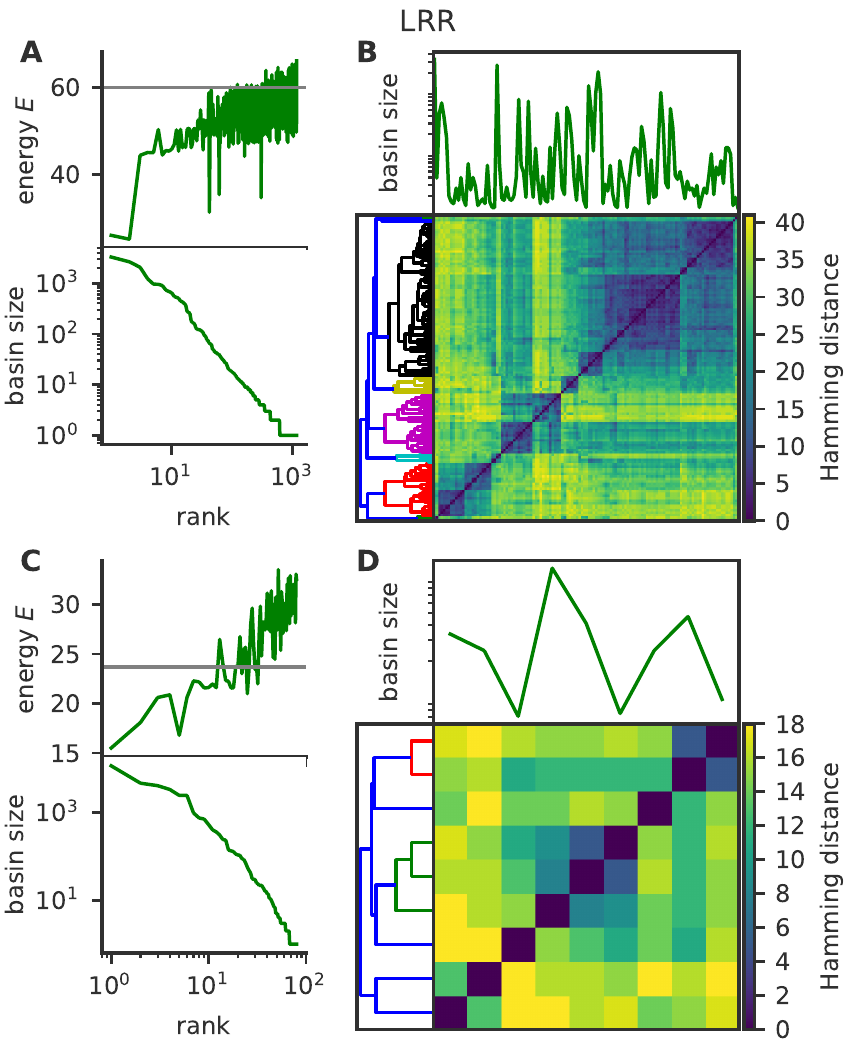}
\caption{
\captionSthree
\label{fig4_lrr:fig}
}
\end{figure}

\begin{figure}
\includegraphics[width=\linewidth]{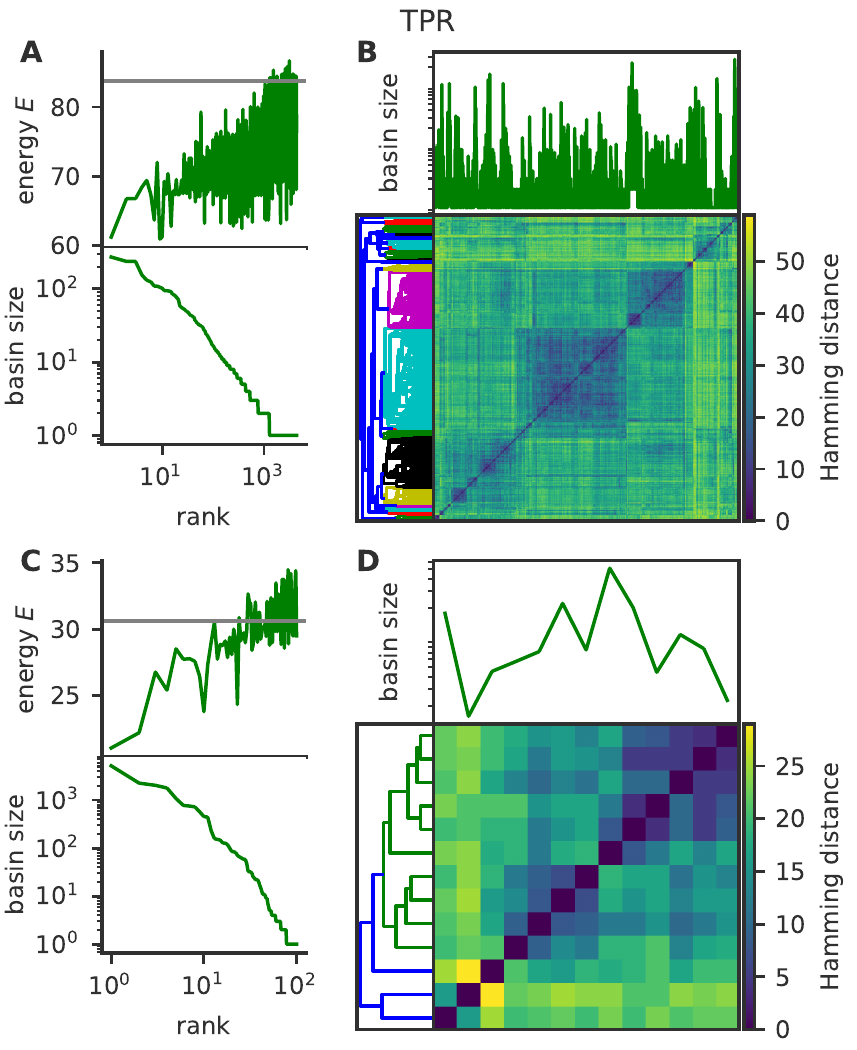}
\caption{
\captionSfour
\label{fig4_tpr:fig}
}
\end{figure}

\begin{figure}
\includegraphics[width=\linewidth]{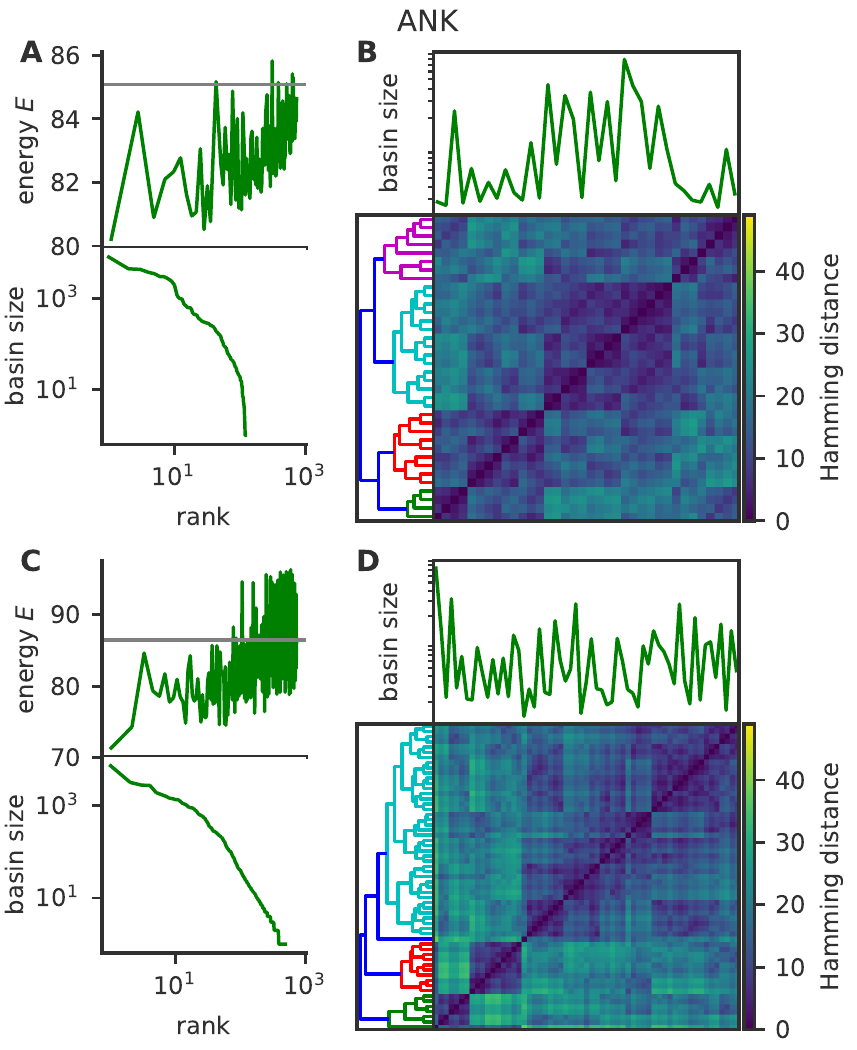}
\caption{
\captionSfive
\label{fig4_W:fig}
}
\end{figure}

\begin{figure}
\includegraphics[width=\linewidth]{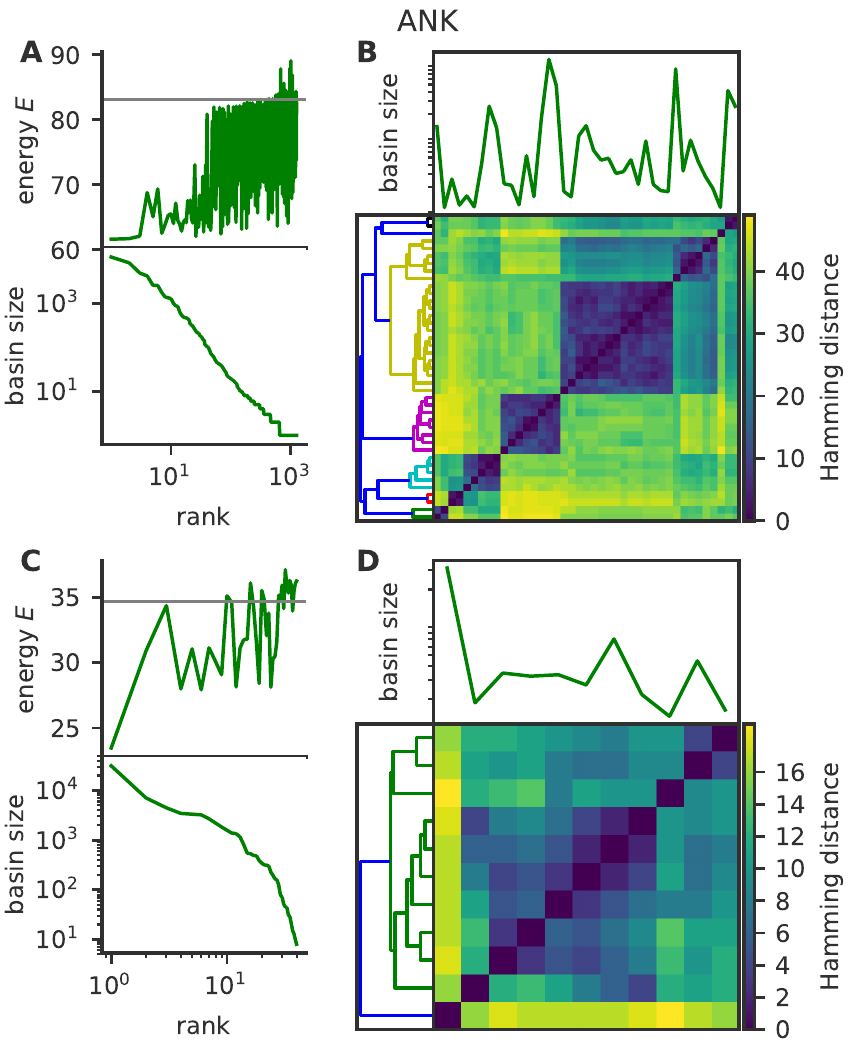}
\caption{
\captionSsix
\label{fig4_synth:fig}
}
\end{figure}

In the end we can repeat the quenching procedure described above for LRR and TPR families. The result are sketched in Fig.~\ref{fig4_lrr:fig}  and Fig.~\ref{fig4_tpr:fig} and lead to similar conclusions as for the ANK family.

\ssection{Kullback-Leibler divergence}~\label{methods:DKL}

The  Kullback-Leibler divergence between two families ${\rm A}$ and ${\rm B}$ is defined as $D_{\rm KL}({\rm A} || {\rm B}) = \sum_{\bs} p_{\rm A}(\bs) \log_2{p_{\rm B}(\bs)/p_{\rm A}(\bs)}$.
We can substitute the sequence ensembles for ANK and TPR in the definition of the probabilities obtaining:
\begin{equation}\label{Dkl_ANK_TPR} 
D_{\rm KL}(ANK || TPR)= \langle E_{\rm TPR} - E_{\rm ANK} \rangle_{\rm ANK} + F_{\rm ANK} - F_{\rm TPR} \,,
\end{equation}
\begin{equation}\label{Dkl_TPR_ANK} 
D_{\rm KL}(TPR || ANK)= \langle E_{\rm ANK} - E_{\rm TPR} \rangle_{\rm TPR} + F_{\rm TPR} - F_{\rm ANK} \,,
\end{equation}
where the notation $ \langle  \rangle_{\rm ANK}$ means that the average is calculated over sequences drawn from the ANK ensemble: $P(\bs)_{\rm ANK}=(1/Z_{\rm ANK})e^{-E(\bs)_{\rm ANK}}$.
Therefore $ \langle E_{\rm TPR} \rangle_{\rm ANK}$ is the average TPR energy function evaluated, via the structural alignment between the two families, on $80000$ sequences generated through a Monte Carlo sampling of the ANK model~\eqref{H_ro} (and analogously for $ \langle E_{\rm ANK} \rangle_{\rm TPR}$). The terms $F_{\rm ANK} $ and $F_{\rm TPR}$ are calculated in the same way as when estimating the entropy through Eqs.~\eqref{F_int},\eqref{F_0}, as explained in Section~\ref{entropy_calc_meth:sec}.

For the control against a random polypeptide of length $L$ we use $ D_{\rm KL}({\rm FAM} || {\rm rand})=\log{\Lambda} - S({\rm FAM}) $, where $\Lambda=21^L$ is the total number of possible sequences of length $L$.

{\bf Acknowledgements. 
} This work was partly supported by ERC CoG 724208 and ECOS Sud - MINCyT no. A14E04
\medskip

\bibliographystyle{pnas}

\end{document}